1    **Generalizability of Artificial Neural Network Models**

2    **in Ecological Appplications:**

3    **Predicting Nest Occurrence and Breeding Success**

4    **of the Red-winged Blackbird *Agelaius phoeniceus***

7    Uygar Özesmi[a], Can O. Tan[b] , Stacy L. Özesmi[a]*, and Raleigh J. Robertson[c]

9    [a]Environmental Science Chair, Department of Environmental Engineering, Erciyes

10   University, 38039 Kayseri Turkey

11   [b]Department of Biology, Middle East Technical University, 06531 Ankara Turkey

12   [c]Department of Biology, Queen's University, Kingston, Ontario Canada

13   * Corresponding author. Environmental Science Chair, Department of Environmental

14   Engineering, Erciyes University, 38039 Kayseri Turkey

15   Tel: 90 352 437 6748 Fax: 90 352 437 4404 email: stacy@erciyes.edu.tr

17   **Abstract**

18          Separate artificial neural network (ANN) models were developed from data in

19   two geographical regions and years apart for a marsh-nesting bird, the red-winged

20   blackbird *Agelaius phoeniceus*. Each model was independently tested on the spatially and

21   temporally distinct data from the other region to determine how generalizable it was.  The

22   first model was developed to predict occurrence of nests in two wetlands on Lake Erie,

23   Ohio in 1995 and 1996. The second model was developed to predict breeding success in

24    two marshes in Connecticut, USA in 1969 and 1970.  Independent variables were

25    vegetation durability, stem density, stem/nest height, distance to open water, distance to

26    edge, and water depth.  The nest occurrence model performance on the training data was

27    at an average cross entropy, or concordance index (c-index), of 0.75. Within geographical

28    region testing in two different wetlands resulted in c-indices of 0.66 and 0.53. The

29    breeding success model performance was at a c-index of 0.75 on the training data and at

30    c-indices of 0.47 and 0.53 for within region testing. When we tested the nest occurrence

31    model on fledged nestling data we obtained c-indices of 0.69 and 0.47 in Clarkes Pond in

32    1969 and 1970 respectively, and 0.43 and 0.52 in All Saints Marsh in 1969 and 1970

33    respectively. When we tested the fledged nestling model on the nest occurrence data, we

34    obtained c-indices of 0.70 and 0.41 in Stubble Patch in 1995 and 1996 respectively, and

35    0.54 and 0.55 for Darr in 1995 and 1996 respectively. With input variable relevances,

36    sensitivity analyses and neural interpretation diagrams we were able to understand how

37    the different models predicted nest occurrence and breeding success and compare their

38    differences and similarities.  Important variables for predicting nest site

39    selection/breeding success in both models were vegetation durability and distance to open

40    water.  Both models also predicted increasing nest occurrence/breeding success with

41    increasing water depth under the nest and increasing distance to edge.  However,

42    relationships for prediction differed in the models.  Generalizability of the models was

43    poor except when the marshes had similar values of important variables in the model, for

44    example water depth.  ANN models performed better than generalized linear models

45    (GLM) on marshes with similar structures.  Generalizability of the models did not differ

46    in nest occurrence and breeding success data. Extensive testing also showed that the

47  GLMs were not necessarily more generalizable than ANNs.  The results from this study

48  suggest that ANN models make good definitions of a study system but are too specific to

49  generalize well to other ecologically complex systems unless input variable distributions

50  are very similar.

51

52  *Keywords*: Habitat models; General linear model; Spatial habitat selection; Habitat

53  preference; Freshwater marsh; Wetlands; Model validation.

54

55

56

57  **1.  Introduction**

58

59         Understanding bird habitat selection and predicting presence, absence or

60  breeding success of a species in particular locations under certain environmental

61  conditions in time and space is one of the key issues for wildlife management and

62  conservation. The difficulty arises as most of the ecological processes and

63  relationships, if not all, exhibit nonlinearity and are very complex in nature.

64  However, artificial neural networks (ANNs) have been shown to be a successful tool

65  for interpretation and prediction under these conditions. The aim of this study is to

66  test the generalizability of spatial habitat models of red-winged blackbirds (*Agelaius*

67  *phoeniceus*) and to observe the commonalities between models to find out

68  generalizable processes. Finally we examine whether nest occurrence models could

69  predict breeding success and vice versa.

70    ANN models have been applied to many diverse ecological problems from

71    phytoplankton production (Scardi, 1996, 2001) to community changes based on

72    climatic inputs (Tan and Smeins, 1996) to animal damage to crops (Tourenq *et al.*,

73    1999; Spitz and Lek, 1999). ANNs have been used to study the relationship between

74    species and habitat variables for birds (Fielding, 1999b; Manel *et al.*,1999),

75    cyanobacteria (Maier *et al.*, 1998), fish (Baran *et al.*, 1996; Reyjol *et al.*, 2001), and

76    macroinvertebrates (Hoang *et al.*, 2001).  However one goal for the use of ANNs in

77    ecology, which has been elusive, has been the development of generalizable models

78    that can be applied to different systems in different geographical regions.  This is

79    currently an active area of research.  For example, Wilson and Recknagel (2001)

80    developed generic ANN lake models to predict algal abundances in freshwater lakes.

81    Scardi (2001) explained the use of a metamodeling procedure to generalize neural

82    network models to other geographic regions when a large amount of data is not

83    available for those regions.

84        In order to assess whether we have achieved our goal of generalizability it is

85    necessary to have an independent test of the model (Fielding, 1999a).  A model that has

86    not been tested is only a definition of a system.  It becomes a scientific pursuit, a

87    hypothesis, when one starts testing it. An independent test turns a model into a hypothesis

88    that can then be validated (Ayer, 1936). Hirzel *et al.* (2001) argued that comparing

89    models of habitat suitability on one unique case is not enough.  Therefore they created a

90    virtual species and different scenarios to test their models.  We agree that more than one

91    independent test of a model is needed.  For this reason, in this study we tested two ANN

92    spatial habitat models for the red-winged blackbird, *Agelaius phoeniceus*, on data from

93    marshes in different geographical areas and years apart.  This gave four completely

94    independent cases with which to test the generalizability of each model.   We also

95    examine the input variable relevances, neural interpretation diagrams, and sensitivity

96    analyses of both models to understand how they are making predictions.  Finally we

97    make multimodel inferences to suggest how to create a more general model.

98         One of our models was created to predict breeding success based on habitat

99    variables.  The other model was originally developed to predict nest occurrence based on

100   habitat variables.  When developing the nest occurrence model it was assumed that the

101   red-winged blackbirds selected nest locations in places with the highest habitat suitability

102   or quality and that by selecting this optimal habitat they would have high breeding

103   success.  However other limiting factors such as intraspecific competition, interspecific

104   competition, and dispersal capability also influence habitat selection (Cody, 1981;

105   Burger, 1985; Cody, 1985).  In addition some research has questioned the relationship of

106   habitat variables to breeding success.  For example, in a study of savannah sparrow

107   *Passerculus sandwichensis*, Bedard and Lapointe (1983) found that reproductive success

108   depended on predators and catastrophic events such as flooding.  Van Horne (1983)

109   stressed that density must be coupled with demographic data or some erroneous

110   conclusions on what constitutes optimal habitat could result.  Nevertheless, many studies

111   on red-winged blackbirds have found a relationship between reproductive success and

112   habitat variables such as water depth, vegetation type and durability, nest height, and nest

113   location (i.e. Meanley and Webb, 1963; Goddard and Board, 1967; Holcomb and Twiest,

114   1968; Francis, 1971; Robertson, 1972; Holm, 1973; Caccamise, 1977; Weatherhead and

115   Robertson, 1977).  Therefore we assumed in this research that when our model predicts a

116 high probability of nest occurrence, this corresponds to a location with high breeding

117 success and vice versa.

118

119 **2. Methods**

120

121 *2.1 Study areas*

122

123   The study areas consisted of four marshes, two located in Sandusky Bay on Lake

124 Erie, Ohio, USA and two located in Connecticut, in the northeastern USA.  The data were

125 from the years 1995 and 1996 and from the years 1969 and 1970 respectively.

126

127   *2.1.1 Lake Erie marshes*

128

129   The Lake Erie marshes were Stubble Patch and Darr.  Data from these marshes,

130 which included habitat variables and nest occurrence, was collected in 1995 and 1996

131 (Özesmi, 1996).  These two marshes lie in the southwestern end of Lake Erie on

132 Sandusky Bay and are part of Winous Point Wetlands, which consists of many wetland

133 units divided by dikes.  Stubble Patch was a 16.7 ha wetland that had been managed as a

134 shallow emergent marsh since 1990.  In 1995 it was dominated by *Typha angustifolia*,

135 *Phalaris arundinacea*, and *Hibiscus palustris*.  In 1996 *Hibiscus palustris* declined and

136 *Sparganium eurycarpum* dominated large areas of the marsh as did *Phalaris arundinacea*

137 and *Typha angustifolia*.  The average water level for both years was about 40 cm in the

138 vegetated areas.  In 1995 the area with vegetation cover was 4.5 ha but in 1996 vegetation

139    cover increased to 7.3 ha. Open water covered more than half of the marsh in both years

140    (Özesmi, 1996).

141        Darr was a 25.8 ha wetland dominated by *Typha angustifolia* and *Hibiscus*

142    *palustris*. About one-half of that area was open water. The average water level in 1995

143    was 30.6 cm in the vegetated areas. However in 1996 Darr was managed as a shallow

144    marsh and the average water level was 15.3 cm in the vegetated areas. From 1995 to

145    1996 the area of vegetation cover increased form 12.3 ha to 13.4 ha (Özesmi, 1996).

146        In 1995 a total of 85 red-winged blackbird nests were found in Darr and in 1996

147    42 red-winged blackbird nests were found. Darr had 5 marsh wren nests in 1995 but 28

148    marsh wren nests in 1996. Competition with marsh wren dramatically affected red-

149    winged blackbird habitat selection as the marsh wren occupied the most central and

150    deepest areas of the marsh (Özesmi and Özesmi, 1999). In Stubble Patch 30 and 49 red-

151    winged blackbird nests were found in 1995 and 1996 respectively. Stubble Patch did not

152    have any marsh wren nests. Thus the Lake Erie wetlands were much larger in size with a

153    much lower density of red-winged blackbird nests than the Connecticut wetlands (Table

154    1).

155        In addition to the nest occurrence data, there was some information on breeding

156    success available for Darr and Stubble Patch in 1996, although the nests were not

157    followed until all the eggs were hatched and fledged. If at the time of the last nest check,

158    the nests had eggs, or fledglings, they were scored as successful. Nests that were empty,

159    showed signs of predation, or were knocked down were scored as unsuccessful.

160

161    *2.1.2 Connecticut marshes*

162

163     The two marshes in Connecticut were Clarkes Pond and All Saints Marsh.  The

164     data from these marshes, which included habitat information and breeding success, was

165     collected in the years 1969 and 1970 (Robertson, 1972).  Clarkes Pond was a constructed

166     impoundment 4.65 ha in size on the Mill River in Hamden, Connecticut.  The dominant

167     emergent vegetation, which covered 1.92 ha of the marsh, was *Typha latifolia* and *Typha*

168     *angustifolia.*  Average depth of the water underneath nests in the emergent vegetation

169     was about 42 cm.  The remaining area, 2.73 ha, was open water with some water lily

170     (*Nymphea*), pickerelweed (*Pontederia*) and arrowhead (*Sagittaria*).  The surroundings of

171     Clarkes Pond consisted of large areas of mixed deciduous woodlands, a pine plantation, a

172     pasture, and a mowed field. Clarkes Pond had 202 and 167 red-winged blackbird nests in

173     1969 and 1970 respectively (Robertson, 1972).

174     All Saints Marsh was located northeast of New Haven, Connecticut about 15 km

175     away from Clarkes Pond.  Its area was 1.09 ha and it was covered with emergent

176     vegetation consisting of *Typha latifolia* interspersed with small patches of open water.

177     Dense areas of buttonbush (*Cephalanthus occidentalis*) as well as other bushes were

178     scattered throughout the marsh, which were often used for nesting.  The marsh seemed to

179     be spring fed with an average water depth of 30-40 cm during the nesting season.  All

180     Saints Marsh was surrounded by extensive areas of mixed deciduous woodland on two

181     sides and weedy fields in early stages of old field succession on two sides.  All Saints

182     Marsh had 108 red-winged blackbird nests in 1969 and 128 in 1970 (Robertson, 1972).

183

184     *2.2 Artificial neural network analysis*

185

186　　　　The ANNs used were the feed-forward multi-layer perceptron with back-

187　propagation for training, which were created using NevProp3 software (Goodman, 1996).

188

189　　　　2.2.1 Lake Erie ANN nest occurrence model

190

191　　　　The nest occurrence model was trained using 1995 data from the Lake Erie

192　wetlands Stubble Patch and Darr.  The independent variables were vegetation durability

193　based on an ordinate scale between 0 and 100 (Özesmi and Mitsch, 1997), stem density

194　(number of stems/m$^2$), stem height (cm above water), distance to open water (m),

195　distance to edge (m), and water depth (cm). All inputs were standardized to a mean of

196　zero with the units in standard deviations. The dependent variable was a binary index of

197　nest occurrence.

198　　　　For hidden units, we used a symmetric logistic function ranging from –0.5 to 0.5

199　and for output layer an asymmetric logistic ranging from 0 to 1. The output was a

200　probability of nest occurrence between 0 and 1. The score threshold was set to 0.5, so that

201　probability predictions of the network lower than 0.5 were classified as no nest (0) and

202　above 0.5 as nest (1). The training algorithm used was gradient descent.  The initial

203　values of the random weights were randomly set between ±0.1. We used a learning rate

204　of 0.01. We trained the model using Darr 1995 data.  We used early stopping with

205　Stubble Patch 1995 data as a holdout data set to determine when to stop training the

206　ANN. We tried different configurations of architecture for the model but chose a single

207　hidden layer with six hidden units for the final model because greater numbers of hidden

208 units and hidden layers did not give better results on the training data. The final model

209 was tested on Stubble Patch and Darr data from 1996. More detailed information on how

210 the nest occurrence model was developed can be found in Özesmi and Özesmi (1999).

211

212 *2.2.2 Clarkes Pond ANN breeding success model*

213

214 In the breeding success model the independent variables were vegetation

215 durability based on an ordinate scale between 0 and 100 (Özesmi and Mitsch, 1997),

216 nest height (cm), distance to open water (m), distance to edge (m) and water depth

217 (cm). We standardized all inputs to a mean of zero with the units in standard

218 deviation. A binary index of whether or not any nestlings fledged was the dependent

219 variable.

220 For hidden units, we used a symmetric logistic function ranging from –0.5 to

221 0.5 and for output layer an asymmetric logistic ranging from 0 to 1. Thus the output

222 was a probability of fledglings between 0 and 1. The score threshold was set to 0.5, so

223 that probability predictions of the network lower than 0.5 were classified as not

224 fledged (0) and above 0.5 as fledged (1).

225 *2.2.2.1 Training the model*

226 When we varied the training algorithm for the networks we found that the

227 corrected c-index was consistently higher for QuickProp than with pure gradient

228 descent. Therefore we used the Quickprop algorithm, which is also faster than pure

229 gradient descent. Quickprop increases the learn rate for each weight's dimension if

230 the global error decreases and assumes that the error surface curvature is locally

231    parabolic (quadratic), so that each change in the gradient and weight uniquely

232    determines the bottom of the current parabola (Goodman, 1996). Quickprop tries to

233    move towards that minimum.

234    We developed the model using data from one marsh, Clarkes Pond in 1969-

235    1970. In total 294 data points were available to train the model. We used

236    bootstrapping (150 with 66% of data) of the training data to determine the maximum

237    number of epochs to train the model. We stopped the training when corrected average

238    cross entropy (c-index) levelled off. C-index, also known as concordance index, is the

239    probability that the model will give a higher output probability to a case with an

240    actual output of 1 versus a case with an actual output of 0. If the c-index is 1, the

241    model predicts perfectly and if the c-index is 0.5 the model does not perform better

242    than a random model (Goodman, 1996). C-index also corresponds to the approximate

243    area under the Receiver Operating Characteristic (ROC) curve. In a ROC curve the

244    proportion of true-positive predictions (sensitivity) are plotted against the proportion

245    of false-positive predictions (1 minus specificity) for various score thresholds

246    (Fielding, 1999a).

247    We optimized the network parameters of learn rate, weight range, and

248    momentum. Several networks with different combinations of learn rate (ranging

249    between 0.1 and 0.001) and weight range (between 0.1 and 0.001) were run. After

250    optimizing the network architecture (see below), we again optimized the network

251    parameters. We chose 0.01 as the starting learn rate and 0.001 as the range in which

252    the initial random weights were chosen (± weight range).

253    *2.2.2.2 ANN architecture*

254        Although there are heuristic algorithms to determine the network architecture,

255        we systematically explored architectures by running several different models: 1) a

256        general linear model (GLM), (no hidden layer), 2) transformation only model (each

257        input connected to only one hidden unit), 3) one hidden layer having different

258        numbers of hidden units ranging between 2 and 40, 4) two hidden layer networks

259        having 2, 3, 4 and 5 hidden neurons in each layer.  All networks were run 5 times

260        using the same predetermined random seeds, produced by a random number

261        generator, to see the variation in the model performance due to different initial

262        weights.  All the models with one hidden layer performed better than the general

263        linear model and transformation only model.  The models with two hidden layers did

264        not perform better than the models with one hidden layer, thus one hidden layer was

265        chosen for the final network configuration.  The accuracy of the one hidden layer

266        networks first increased with increasing numbers of hidden units and then levelled off

267        around 20 hidden units.

268        For good generalizability, it is necessary to have about 10 times as many training

269        data points as there are weights in the network (Bishop, 1995).  In our case, we had about

270        300 training input – output cases, so we should have approximately 30 connections in our

271        network.  We had 5 input variables and one output variable so this would mean about 4

272        hidden units arranged in the form of a single layer. Nevertheless, we obtained the best c-

273        index with a network having 20 hidden units (c-index = 0.79), which gave 120

274        connections.  For that reason, we tried to optimize the network with 4 hidden units (in

275        terms of network parameter settings), and compare the model accuracies and sensitivity

276        analyses both with the best network having 20 hidden units (120 connections) and the

277  best network having 4 hidden units (24 connections).  Finally, we chose the network with

278  4 hidden units (c-index = 0.75) for the following reasons.  First, given models with

279  similar errors on the training sets, the simpler model is more likely to predict better on

280  new cases, the dictum of Occam's razor.  More formally a criteria such as Akaike

281  Information criterion (AIC), which consists of an error term with a penalty for model

282  complexity, could be used to choose a model (Burnham and Anderson, 2002; Bishop,

283  1995).  This also would have resulted in choosing the model with 4 hidden units.

284  Because increasing the number of estimated parameters (weights) in a model usually

285  always increases its predictive ability, it is necessary to account for this increased model

286  complexity when deciding which model to choose.  Thus current literature is

287  recommending the use of criteria to select a model when many models are created with

288  the same dataset (Burnham and Anderson, 2002; McNally, 2000).

289      Second, we ran the two models with 4 and 20 hidden units with the same ten

290  random starts to determine the variation in the c-index and corrected c-index (based on

291  150 bootstraps).  For the model with 4 hidden units, the c-indices did not vary as much as

292  with the 20 hidden units model (Table 2).  We thought the larger variation in c-indices of

293  the 20 hidden unit model might be an indication of overfit on the training data.    Third,

294  according to the sensitivity analyses the network with 4 hidden units seemed to capture

295  the relationship between inputs and the output better; that is, it made more sense based on

296  what is known about red-winged blackbird habitat selection.

297      *2.2.2.3 Model test*

298      The final model was tested using the data from All Saints Marsh from 1969

299  (n=101) and from 1970 (n=130).  We compared the performance of the ANN with a

300      GLM, an ANN with no hidden units, which is basically a logistic regression model,

301      on the training data from Clarkes Pond and on the All Saints Marsh test data.

302

303      *2.2.3 Comparison of models*

304

305      In this study the breeding success model, which had been trained on Clarkes Pond

306      1969-1970 data, was tested on the Lake Erie wetlands data from Darr and Stubble Patch

307      in 1995 and 1996.  The Lake Erie nest occurrence model was tested on the Connecticut

308      wetlands data from Clarkes Pond and All Saints in 1969 and 1970. Because one model

309      was developed to predict breeding success and the other nest occurrence, for this research

310      we assumed that a high probability of nest occurrence corresponds to a high probability

311      of breeding success and vice versa (for more discussion on this assumption see the

312      introduction). In addition, since the Connecticut wetland variables did not include stem

313      density, the average value of stem density from the Lake Erie wetlands was used when

314      testing Connecticut wetlands data on the Lake Erie model.  Note also that nest height was

315      available for the Connecticut wetlands but that stem height was available for the Lake

316      Erie wetlands.  Stem heights were about 50 cm higher than nest heights on average.  The

317      assumption was made that stem height and nest height were correlated.  Thus when the

318      models were tested, nest heights were input instead of stem heights in the nest occurrence

319      model and vice versa for the breeding success model.

320      In addition to the nest occurrence data from Darr and Stubble Patch, the

321      information on breeding success from Darr and Stubble Patch in 1996 together with the

322      habitat variables of vegetation durability, nest height, distance to open water, distance to

323  edge, and water depth under the nest was used as test data for the Clarkes Pond breeding

324  success model.

325  Input variable relevances and neural interpretation diagrams (NIDs) (Özesmi

326  and Özesmi, 1999), as well as sensitivity analyses (Lek *et al.*, 1996; Scardi, 1996;

327  Recknagel *et al.*, 1997) were used to understand the models' predictions.  The input

328  variable relevance is the sum square of weights for one input variable divided by the

329  sum square of weights for all input variables and thus it shows the contribution of a

330  variable to the model.  In a NID, the weights of the connections between units are

331  represented by pixel weights of lines scaled to the relative values of the weights in the

332  ANN.  Black lines represent positive signals and gray lines represent negative signals.

333  By looking at the weights and signs of the connections in a NID it is possible to see the

334  importance of variables in the model and the interactions between variables. Sensitivity

335  analyses are done by varying the values of one input variable at a time while holding

336  the other input variables at constant values (we used average values) and plotting the

337  values against probability of breeding success.

338

339  **3. Results and discussion**

340

341  *3.1 Within geographic region model results*

342

343  *3.1.1 Lake Erie ANN and GLM nest occurrence model*

344

345      The Lake Erie ANN nest occurrence model performance on the training data,

346      which was Stubble Patch and Darr data for 1995, was at a c-index of 0.75 (Table 3).

347      When the model was tested on the 1996 data for Stubble Patch and Darr the resulting c-

348      indices were 0.66 and 0.53 respectively. C-indices of the Lake Erie GLM model on the

349      test data were 0.67 and 0.57 for Stubble Patch and Darr respectively.  The Lake Erie

350      ANN nest occurrence model did not perform well on Darr in 1996 because of the

351      presence of marsh wren.  More details on the Lake Erie models as well as a marsh wren

352      model and a marsh wren and red-winged blackbird interaction model can be found in

353      Özesmi and Özesmi (1999).

354

355      *3.1.2 Clarkes Pond ANN and GLM breeding success model*

356

357      The Clarkes Pond ANN model of breeding success, which was defined as at least

358      one nestling fledged, had a c-index of 0.75 for the training data of Clarkes Pond in 1969

359      and 1970 (Table 3).  The corrected c-index, calculated using 150 bootstraps, levelled off

360      at 0.66 after 70 epochs. This corrected c-index should reflect how generalizable the

361      model is.

362      The Clarkes Pond breeding success ANN model was tested independently on

363      All Saints Marsh.  All Saints Marsh was quite different from Clarkes Pond (Table 1).

364      Clarkes Pond was an open water pond surrounded by *Typha* vegetation where the red-

365      winged blackbirds nested.  All Saints Marsh was a shallow emergent marsh

366      dominated by *Typha* with scattered shrubs throughout.  All Saints Marsh had a lower

367      water depth on average than Clarkes Pond.  In addition, All Saints Marsh was about

368    one-fourth of the size of Clarkes Pond.  All Saints Marsh did not have any large areas

369    of open water in either 1969 or 1970.  Therefore the distance to open water variable

370    for All Saints Marsh was input in two different ways: 1) as the greatest distance in

371    Clarkes Pond (112 m) and 2) as the mean distance to open water in the Clarkes Pond

372    data.  These two different ways to input distance to open water in All Saints Marsh

373    gave similar results.  A separate model developed without the distance to open water

374    variable as input did not give better results as well.

375        In addition to these structural differences, the density of nests was greater in

376    All Saints Marsh than Clarkes Pond, 108.3 versus 96.1 nests/ha per year.  The

377    percentage of successful nests was also greater in All Saints Marsh for both years,

378    64.4% versus 47.6% average for both years (Robertson, 1972).

379        When this model was tested on All Saints Marsh for 1969 and 1970, the

380    within region testing resulted in c-indices of 0.47 and 0.53 respectively.  The Clarkes

381    Pond GLM model had a c-index of 0.63 on the training data and c-indices of 0.53 and

382    0.53 for All Saints Marsh in 1969 and 1970 respectively.  Thus both the artificial

383    neural network model and general linear model both did not perform better than a

384    random model.

385        One reason for the poor model performance is probably the structural

386    differences of the marshes (different vegetation types, different vegetation/open water

387    ratios, different water depths, different sizes).  In addition, in these marshes predation

388    was a significant factor in breeding success. In particular, Clarkes Pond was under

389    very high predation pressure with 45% and 47.3% of the nests predated in 1969 and

390    1970 respectively (Robertson, 1972).  All Saints Marsh had 11.1% and 16.4 % of the

391    nests predated in 1969 and 1970 respectively.  Racoon (*Procyon lotor*) was

392    responsible for the most nest predations, accounting for 54 of 80 nests predated in the

393    nestling phase in these marshes. Other nest predators were likely birds or water

394    snakes.  Long-billed marsh wrens (*Cistothorus palustris*) were present in Clarkes

395    Pond in 1969 but there was no evidence that these birds destroyed red-winged

396    blackbird nests.

397        Predation seemed to dominate over other environmental factors (the habitat

398    input variables) for determining breeding success.  However, environmental variables

399    affect predation.  It is more difficult for a predator to get to a nest in deeper water, in

400    the center of a marsh, and surrounded by open water.  For example, predation

401    decreased with increasing water depth under the nest.  Although predators are

402    influenced by environmental variables included in the model, complete reliance on

403    environmental variables to account for predation is not sufficient and the inclusion of

404    predators into the model can improve performance (Özesmi and Özesmi, 1999).

405

406    *3.2 Model generalizability test results*

407

408        *3.2.1 Lake Erie nest occurrence ANN and GLM models*

409

410        The Lake Erie ANN nest occurrence model tested on the Connecticut breeding

411    success data resulted in c-indices of 0.69 and 0.47 for Clarkes Pond in 1969 and 1970

412    respectively, and 0.43 and 0.52 for All Saints Marsh in 1969 and 1970 respectively

413    (Table 3).  Thus the Lake Erie ANN nest occurrence model performed well on Clarkes

414    Pond 1969 data but no better than a random model on the other data. Clarkes Pond 1969

415    data were the most similar to the training data for the Lake Erie model, showing closer

416    distances to open water and deeper water for higher breeding success (Figures 1a-b).

417    These variables had high relevances in the Lake Erie ANN model.

418        The Lake Erie GLM model tested on Clarkes Pond resulted in c-indices of 0.66

419    and 0.48 for 1969 and 1970 respectively.  All Saints Marsh in 1969 and 1970 had c-

420    indices of 0.38 and 0.48 respectively.  The GLM model performed very poorly on all

421    marshes and years except Clarkes Pond in 1969.  This model relied heavily on distance to

422    open water (Table 4).  Similar to the training data, Darr and Stubble Patch in 1995, which

423    had higher nest occurrence close to open water, Clarkes Pond in 1969 had higher

424    breeding success for closer distances to open water (Figure 1a).  However, Clarkes Pond

425    in 1970 showed the opposite and All Saints Marsh did not have large areas of open water

426    in either year.

427

428        *3.2.2 Clarkes Pond breeding success ANN and GLM models*

429

430        The Clarkes Pond ANN breeding success model tested on the Lake Erie nest

431    occurrence data gave c-indices of 0.70 and 0.41 for Stubble Patch in 1995 and 1996

432    respectively (Table 3).  For Darr the c-indices were 0.54 and 0.55 in 1995 and 1996

433    respectively.  The Clarkes Pond ANN breeding success model performed well on Stubble

434    Patch 1995 data but no better than a random model on the other data.  One reason for the

435    good performance on Stubble Patch in 1995 may have been because in this year Stubble

436    Patch had the most open water and thus would have been most similar to Clarkes Pond

437 (Table 1) in terms of vegetation/open water ratio. The mixture of vegetation cover and

438 open water is known to be an important factor in red-winged blackbird nest site selection

439 (Orians, 1980).

440     The Clarkes Pond ANN breeding success model and the Lake Erie nest

441 occurrence ANN model are predicting differently. The Lake Erie model predicts high

442 nesting probability in areas near to open water and in the center of the marsh while the

443 Clarkes Pond model probabilities seem to be based on water depth together with other

444 variables.

445     When the Clarkes Pond ANN breeding success model was tested on Darr and

446 Stubble Patch breeding success data from 1996 c-indices of 0.64 and 0.52 respectively

447 were obtained (Table 3).

448     The Clarkes Pond GLM breeding success model tested on the Lake Erie nest

449 occurrence data resulted in c-indices of 0.60 and 0.60 on Darr in 1995 and 1996, and 0.62

450 and 0.71 on Stubble Patch in 1995 and 1996 (Table 3). Thus the Clarkes Pond GLM

451 breeding success model performed better than the ANN model, except for on Stubble

452 Patch in 1995. This GLM model relied heavily on water depth to predict breeding

453 success (Table 4). The water depth of Clarkes Pond was most similar to Stubble Patch,

454 especially in 1996 (Figure 1b).

455     The Clarkes Pond GLM breeding success model tested on Darr and Stubble Patch

456 1996 breeding success data gave c-indices of 0.57 and 0.57 respectively. Next we made

457 informal multimodel inferences using input variable relevances, sensitivity analses, and

458 neural interpretation diagrams.

459

 *3.3 Input Variable Relevances*

461

462      Input variable relevances for both models are shown in Table 4.  The variables

463 with the highest relevances for the Lake Erie ANN nest occurrence model were distance

464 to open water, vegetation durability, stem density and distance to edge (Özesmi and

465 Özesmi, 1999).  For the Clarkes Pond breeding success model, the highest input variable

466 relevances were nest height, water depth, distance to open water, and vegetation

467 durability.  Both models are in agreement that two of the four most important variables

468 for red-winged blackbird nesting and fledging probability are distance to open water and

469 vegetation durability.  The importance of vegetation structure for red-winged blackbirds

470 has previously been noted (Goddard and Board, 1967; Holm, 1973; Burger, 1985).

471 Murkin *et al.* (1989) noted that red-winged blackbirds seemed to prefer a certain distance

472 to open water for nest placement.

473      According to the average values for distance to open water, and also water depth,

474 the Lake Erie model data (Stubble Patch and Darr 1995) were most similar to Clarkes

475 Pond 1969 data (Figures 1a-b).  Distance to open water had a high relevance in the Lake

476 Erie model.  Another reason the nest occurrence model may have performed better on

477 Clarkes Pond 1969 data is that marsh wrens were present in that year but not in 1970 or

478 in All Saints Marsh (Table 1).  Five marsh wrens were also present in Darr in 1995.  Darr

479 and Stubble Patch 1995 data were used to train the Lake Erie ANN nest occurrence

480 model.

481      According to mean values for water depth, Clarkes Pond data from 1969 and

482 1970, which were used to develop the Clarkes Pond breeding success model, were closest

483    to the Stubble Patch 1995 data.  As water depth was one of the two most relevant

484    variables in the Clarkes Pond breeding success, this may be one reason why the Clarkes

485    Pond breeding success model performed the best on Stubble Patch 1995 data.

486

487    *3.4 Sensitivity analysis*

488

489    For the Lake Erie nest occurrence model, nesting probability increased with

490    increasing vegetation durability, increasing water depth, and increasing distance to edge

491    (Özesmi and Özesmi, 1999).  Nesting probability increased with increasing stem height

492    to about 130 cm and then decreased.  Nesting probability increased with increasing

493    stem density to about 160 stems/m$^2$ and then decreased.  Nesting probability decreased

494    with increasing distance to open water.

495    For the Clarkes Pond breeding success model, sensitivity analyses indicated that

496    breeding success increased with increasing water depth and increasing distance to edge

497    (Figure 2a-e).  Breeding success generally decreased with increasing nest height and

498    increasing distance to open water.  Breeding success was high with lower vegetation

499    durability and high vegetation durability while vegetation with mid-range durability had

500    low breeding success.

501    Comparing the sensitivity analyses curves, both the Lake Erie nest occurrence

502    model and the Clarkes Pond breeding success model predict increased nesting and

503    breeding success respectively with increasing water depth and increasing distance to

504    edge.  Weatherhead and Robertson (1977), with their weighted linear nest score that

505    significantly correlated with breeding success for a marsh in Ontario, Canada, gave

506     higher scores to nests with deeper water underneath.  Brown and Goertz (1978),

507     Goddard and Board (1967) and Robertson (1972) also report increased breeding

508     success with greater water depth under the nest.

509          Differences in the models are that the Lake Erie nest occurrence model predicts

510     increased nesting with increasing vegetation durability while the Clarkes Pond breeding

511     success model predicts high breeding success in low durability vegetation and high

512     durability vegetation with a decrease in breeding success in middle range of vegetation

513     durability.  The preference of red-winged blackbirds for vegetation of high durability

514     has been noted before (Robertson, 1972; Albers, 1978; Berstein and McLean, 1980;

515     Özesmi and Mitsch; 1997).  These studies are in agreement with the Lake Erie model.

516     However, Weatherhead and Robertson (1977) scored less durable grasses higher than

517     *Typha* but *Typha* higher than sedges for an Ontario, Canada marsh.  This is more in

518     agreement with the Clarkes Pond model.

519          Another difference is that the Lake Erie nest occurrence model predicts an

520     optimum stem height while the Clarkes Pond breeding success model predicts

521     increasing breeding success with decreasing nest height.  Some studies showed that

522     nesting success increased with increasing nest height (Meanley and Webb, 1963;

523     Holcomb and Twiest, 1968) and some with decreasing nest height (Goddard and Board,

524     1967).  The nest score of Weatherhead and Robertson (1977) gave a higher score to

525     nests that were lower, which is in agreement with the Clarkes Pond model.  Brown and

526     Goertz (1978) found that breeding success was the highest in nests 1.2-2.4 m high with

527     lower and higher nests less successful.  Caccamise (1977) showed increasing breeding

528     success with nests up to 1.6-1.79 m high and then decreasing success.  These two

529     studies are more in agreement with the Lake Erie model.  The relationship between nest

530     height and breeding success is not a simple one but may also be related to vegetation

531     type or durability (Francis, 1971) and other environmental factors, such as depth of

532     water under the nest.  Brown and Goertz (1978) noted that red-winged blackbird nests

533     tended to be lower over open water.  In addition, the difference between the Lake Erie

534     model and Clarkes Pond model may also be caused by measurement of different

535     variables, stem height versus nest height.  Stowers *et al.* (1968) reported that red-

536     winged blackbird average nest heights were different in different vegetation types.

537         The third difference is that the Lake Erie model predicts decreasing nest

538     occurrence with increasing distance to open water.  The relationship between breeding

539     success and distance to open water is more complicated in the Clarkes Pond breeding

540     success model but this model also predicts decreasing breeding success when the

541     distance to open water increases beyond about 45 m.

542         Stem density was not an input variable in the Clarkes Pond breeding success

543     model.  The Lake Erie model showed an optimal stem density at about the midrange of

544     the maximum stem density found in these wetlands.  Many authors have stated that red-

545     winged blackbirds prefer dense vegetation (i.e. Albers, 1978; Short, 1985).  However,

546     Holm (1973) found that nests in less dense vegetation fledged more young per nest and

547     Weatherhead and Robertson (1977) found that more open nests were more successful

548     than those in dense cover.  Since sensitivity analysis cannot provide direct

549     interpretation of these complex interactions it was necessary to examine the Neural

550     Interpretation Diagrams of the models.

551

 *3.5 Neural Interpretation Diagrams (NIDs)*

553

554      The NID of the Clarkes Pond breeding success ANN model shows the

555   interaction of variables with each other (Figure 3). Breeding success is higher in nests

556   if the vegetation is not durable but the nest height is high and the water depth is deep

557   (unit 6). Fledging probability decreases with high distance to open water, high nest

558   height and low water depth (7). Breeding success increases if the vegetation is durable

559   and the distance to open water is high (8). If the vegetation durability, nest height,

560   distance to open water, and water depth are all high, fledging probability decreases

561   unless the distance to edge is high (9).

562      The neural interpretation diagrams (NIDs) of the two models are similar in that

563   nesting/breeding success probability always decreases with increasing distance from

564   open water, unless water depth is high (Özesmi and Özesmi, 1999).  Both models also

565   show that increasing vegetation durability increases nest occurrence/breeding success.

566   However in the Clarkes Pond model, fledging probability also increases with low

567   vegetation durability if the nest is high up and in deep water.  Other rules in the models

568   are also different.  For example, in the Lake Erie model, nesting probability increases

569   with increasing stem density and height unless vegetation durability, distance to open

570   water and distance to edge increase.   In the Clarkes Pond model, fledging probability

571   decreases if nest height, vegetation durability, distance to open water, and water depth

572   are all high.  In the Lake Erie model, as distance to open water decreases nesting

573   probability increases but only in areas where stem density and height are enough to

574   support a nest.

575

*3.6 Model generalizability*

576

577

578       These two models agree on some of the important variables and have some of the

579 same relationships to predict nest occurrence and breeding success.  Both models agree

580 on the importance of vegetation durability and distance to open water for nest

581 occurrence/breeding success.  The two models also both predict increased nest

582 occurrence/breeding success with increasing depth of water under the nest and increasing

583 distance to edge.  However, the models are also using different relationships to predict

584 nest occurrence and breeding success.  Therefore the tests of generalizability usually gave

585 poor results.  These results could indicate that either the Lake Erie ANN and GLM nest

586 occurrence model and the Clarkes Pond breeding success model are not generalizable to

587 very different geographic locations and/or marshes of different size and structure, or that

588 nest occurrence is a poor predictor of breeding success

589       Model generalizability was poor except when the marshes had similar values of

590 high relevance input variables, for example water depth.  The artificial neural network

591 models developed were not generalizable to marshes with different sizes and structures.

592 Red-winged blackbirds are highly flexible in their habitat requirements and occupy a

593 wide range of sites from uplands to marshes over a broad geographic area (Orians, 1980).

594 For such an adaptable species, it may be difficult to generalize to different marshes and

595 different geographical regions.  However this research suggests that to create a

596 generalizable model, which can be applied in many different marshes in different

597 locations, we would use the important habitat variables and relationships that are in

598 common for both of these models. Therefore wide testing provides wider understanding.

599     In our research we found that the Clarkes Pond breeding success model predicts

600 nest occurrence as well as the Lake Erie nest occurrence model predicts breeding

601 success (Table 3). However, previous research has indicated that nest occurrence is not

602 always a good predictor of breeding success.  Robertson (1972) concluded that habitat

603 selection by red-winged blackbirds is probably a combination of site tenacity of adults

604 breeding for the second or greater time and selection for the "optimal" breeding habitat

605 by first time breeders.  Thus nests could be located in both optimal and suboptimal

606 habitats.  In addition, Weatherhead and Robertson (1977) found that breeding success

607 was correlated positively with habitat quality and negatively with female breeding

608 density within territories.  Thus females are nesting in territories where their chance of

609 reproductive success is lower than it might have been in another territory.  They

610 postulated that females were cueing in not only on habitat variables but were choosing

611 good mates and thus enhancing their ultimate fitness rather than their immediate

612 reproductive success.  However Holm (1973) asserted that females were only choosing

613 good territories and that was why nesting density was high within some territories.

614 Similarly using the Lake Erie data set we found out that incorporating clustering of

615 nests into a spatial autocorrelation regression model and a spatial data mining tool

616 "Predicting Locations Using Map Similarity (PLUMS)" increased prediction accuracy

617 significantly (Chawla et. al., 2001).

618

619 **4. Conclusion**

620

The ANN models that were tested were very specific to certain conditions of the marshes they were trained on.  However we can expect them to perform well when applied to wetlands with similar values of high relevance input variables, given that there are no other confounding effects, such as the presence of interspecific competitors.  When developing an ANN model, conditions to which the model is applicable should be specified exactly.  Then the model should be tested only on independent data sets that meet those conditions. However, if we want an ANN model that is the most generalizable we need to develop our model on the "average marsh" that has a small number of special cases.  The GLM models developed relied heavily on one input variable and performed better on other wetlands if that variable in those wetlands was also similarly related to nest occurrence/breeding success.  However the extensive testing showed that the GLMs were not necessarily more generalizable than the ANN models.

The ANN models developed are not well suited for predicting and generalizing to other marshes and locations.  However the ANNs are better than the GLMs for defining a system and interpreting the factors that govern nest occurrence/breeding success in a certain location.  In addition, the ANN models developed in different marshes and different locations agree on some of the habitat variables and relationships for predicting nest occurrence/breeding success.  Therefore we suggest a more generalizable model may be created with these variables and relationships.  The process of modelling complex evolutionary and ecological systems humbled us as modellers reminding us that all encompassing relatively simple ecological models, such as ANNs, might not be possible. Yet the modelling

643    process is revealing, allowing us to understand these systems and their peculiarities

644    better. The problem of generalizability in ANN modelling continues to be trapped

645    within the tension of the local and universal in ecology.

646

647    **Acknowledgements**

648

649    We would like to thank Inci Togan for bringing us together to work on modelling and her

650    constant encouragement. Significant computer time was allocated for our research by

651    Erciyes University, Mechanical Engineering Department.

652

764        Tables

765        Table 1.  Comparison of the marshes by size, vegetation, water depth, number of red-

766        winged blackbird (RWB) nests and number of marsh wrens nests.

| Marsh Year Location | Total area (ha) | Vegetated area (ha) | Dominant vegetation | Avg water depth (cm) | RWB Nests | Marsh wrens Nests |
|---|---|---|---|---|---|---|
| Darr 1995 Lake Erie | 25.8 | 12.3 | *Typha, Hibiscus* | 31 | 85 | 5 |
| Darr 1996 Lake Erie | 25.8 | 13.4 | *Typha, Hibiscus* | 15 | 42 | 28 |
| Stubble Patch 1995 Lake Erie | 16.7 | 4.5 | *Typha, Phalaris, Hibiscus* | 40 | 30 | 0 |
| Stubble Patch 1996 Lake Erie | 16.7 | 7.3 | *Sparganium Typha, Phalaris* | 40 | 49 | 0 |
| All Saints 1969 Connecticut | 1.1 | 1.1 | *Typha, bushes* | 30-40 | 202 | 0 |
| All Saints 1970 Connecticut | 1.1 | 1.1 | *Typha, bushes* | 30-40 | 167 | 0 |
| Clarkes Pond 1969 Connecticut | 4.7 | 1.9 | *Typha* | 42 | 108 | some |
| Clarkes Pond 1970 Connecticut | 4.7 | 1.9 | *Typha* | 42 | 128 | 0 |

767

768

768    Table 2.  Variation in model accuracy caused by using 10 different random seeds.

|         | 4 hidden units | | 20 hidden units | |
|---------|---------|------------------|---------|------------------|
|         | c-index | Corrected c-index | c-index | Corrected c-index |
| Mean    | 0.77    | 0.70             | 0.82    | 0.74             |
| Std Dev | 0.01    | 0.01             | 0.03    | 0.03             |
| Maximum | 0.79    | 0.71             | 0.87    | 0.79             |
| Minimum | 0.76    | 0.68             | 0.78    | 0.69             |

769

770

771

771    Table 3.  Model performances (c-index) on training and independent test data sets.

772    Underlined values are model performances on training data.

| | Clarkes Pond GLM Breed. success Model | Clarkes Pond ANN Breed. success Model | Lake Erie GLM Nest occur. Model | Lake Erie ANN Nest occur. Model |
|---|---|---|---|---|
| Clarkes Pond 1969 | <u>0.63</u> | <u>0.75</u> | 0.66 | 0.69 |
| Clarkes Pond 1970 | <u>0.63</u> | <u>0.75</u> | 0.48 | 0.47 |
| All Saints Marsh 1969 | 0.53 | 0.47 | 0.38 | 0.43 |
| All Saints Marsh 1970 | 0.53 | 0.53 | 0.48 | 0.52 |
| Darr 1995 | 0.60 | 0.54 | <u>0.71</u> | <u>0.75</u> |
| Darr 1996 | | | | |
| Nest occurrence | 0.60 | 0.55 | 0.57 | 0.53 |
| Breeding success | 0.57 | 0.64 | | |
| Stubble Patch 1995 | 0.62 | 0.70 | <u>0.71</u> | <u>0.75</u> |
| Stubble Patch 1996 | | | | |
| Nest occurrence | 0.71 | 0.41 | 0.67 | 0.67 |
| Breeding Success | 0.57 | 0.52 | | |
| Average  on test data | 0.59 | 0.55 | 0.54 | 0.55 |
| Min on test data | 0.53 | 0.41 | 0.38 | 0.43 |
| Max on test data | 0.71 | 0.70 | 0.67 | 0.69 |

773

774

774   Table 4. Relevances of input variables for the Clarkes Pond breeding success and the

775   Lake Erie nest occurrence GLM and ANN models.

| Independent Input Variable | Clarkes Pond GLM Model | Clarkes Pond ANN Model | Lake Erie GLM Model | Lake Erie ANN Model |
|---|---|---|---|---|
| Vegetation Durability | 5.8% | 11.2% | 6.4% | 19.2% |
| Stem Density | _____ | _____ | 0.7% | 13.5% |
| Nest Height/Stem Height | 9.9% | 48.2% | 0.9% | 4.6% |
| Distance to Open Water | 0.2% | 13.9% | 88.2% | 45.1% |
| Distance to Edge | 6.5% | 2.1% | 2.9% | 11.0% |
| Water Depth | 77.6% | 24.7% | 1.0% | 6.5% |

776

777

778

**a).**

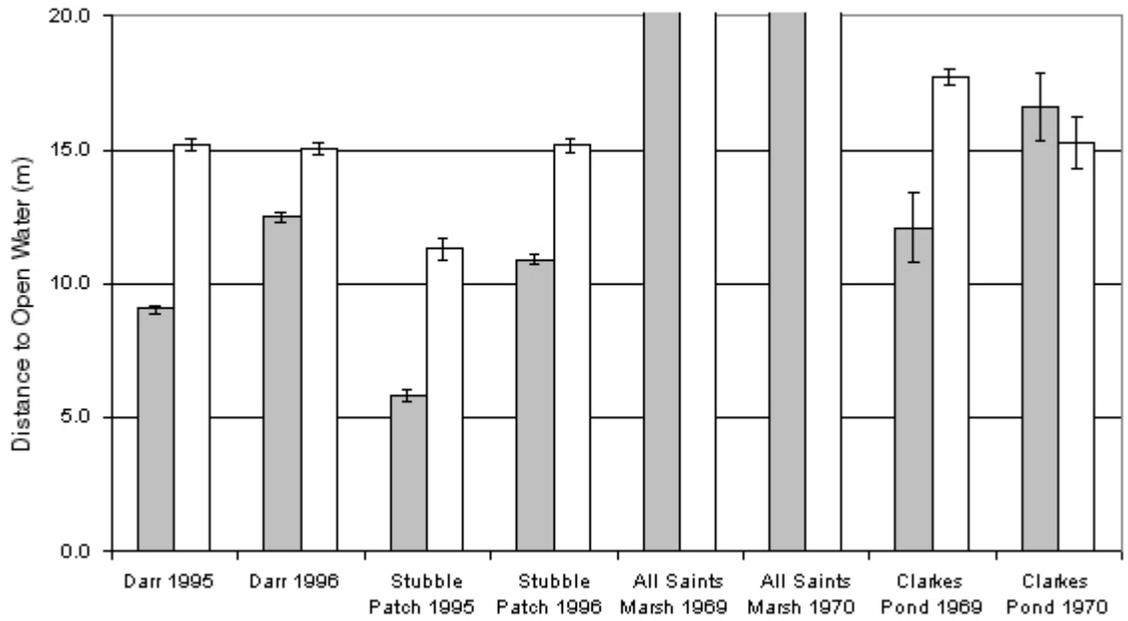

778

**b).**

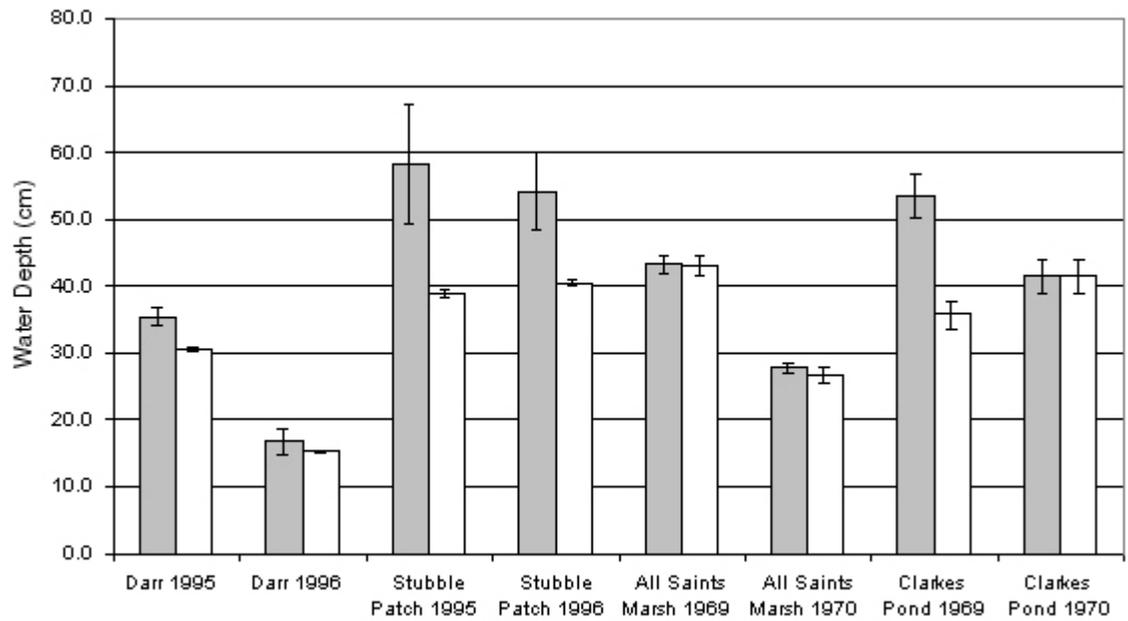

779
780    Figure 1a-b

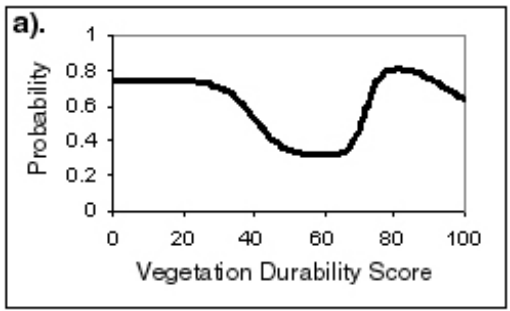

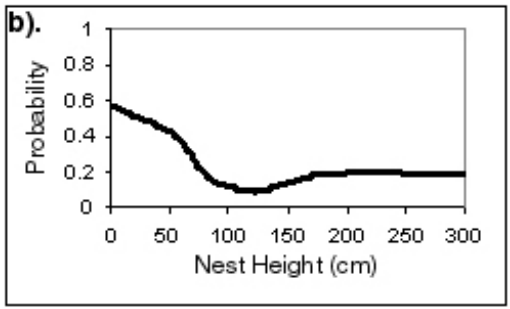

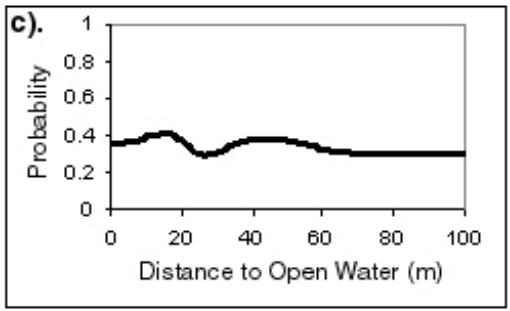

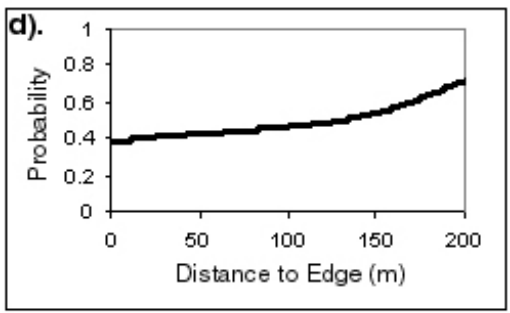

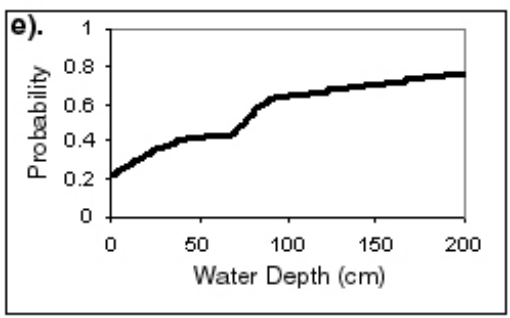

781
782    Figure 2a-e
783

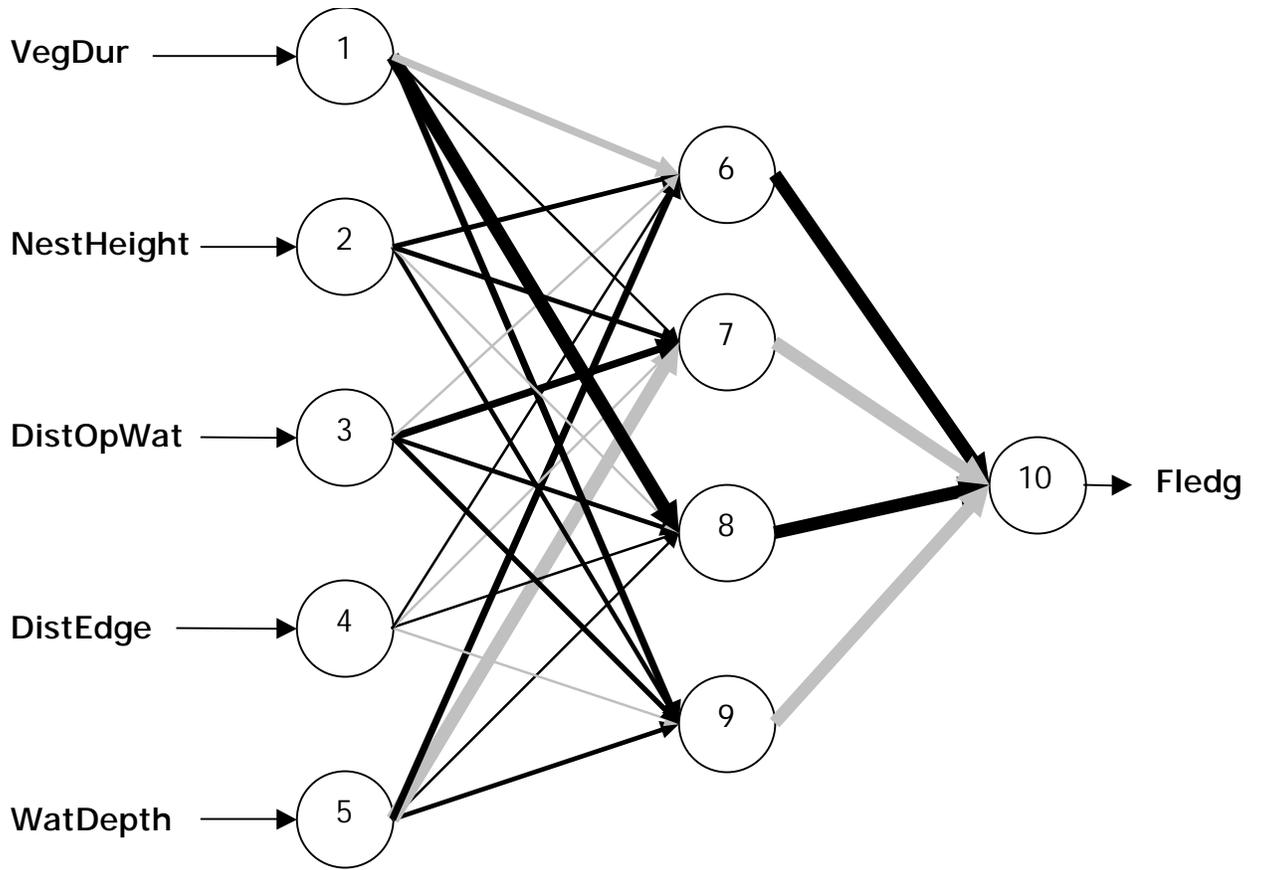

     Figure Captions

786     Figure 1a-b.   Mean and standard errors of distance to open water (m) and water depth

787     (cm) for the different marshes.  Shaded bars indicate nests/fledged, white bars indicate no

788     nests/not fledged.  In figure (a) distance to open water at All Saints Marsh was set to a

789     very high number since there was no open water during 1969-70.

790     Figure 2a-e.  Sensitivity analyses for the Clarkes Pond 1969-70 ANN breeding success

791     model.

792     Figures 3.  Neural Interpretation Diagram (NID) for Clarkes Pond 1969-70 ANN

793     breeding success model. The thickness of each connection is proportional to its

794     relevance.  Black lines represent positive relationships and gray lines represent negative

795     relationships.